\documentclass[epj]{svjour}
\usepackage{graphics}
\newcommand{\be}{\begin{equation}}
\newcommand{\ee}{\end{equation}}
\newcommand{\beq}{\begin{eqnarray}}
\newcommand{\eeq}{\end{eqnarray}}
\def\fq{{f^q}}

\begin{document}
\title{Non-extensive statistical effects in high-energy collisions}
\author{W.M. Alberico\inst{1,3} \and A. Lavagno\inst{2,3}
}                     
\institute{Dipartimento di Fisica, Universit\`a di Torino, Via P.
Giuria 1, I-10124 Torino, Italy \and Dipartimento di Fisica,
Politecnico di Torino, C.so Duca degli Abruzzi 24, I-10129 Torino,
Italy \and Istituto Nazionale di Fisica Nucleare (INFN), Sezione di
Torino, Italy}
\date{Received: date / Revised version: date}
%
\abstract{ Following the basic prescriptions of the relativistic
Tsallis' non-extensive thermostatistics, we investigate from a
phenomenological point of view the relevance of non-extensive
statistical effects on relativistic heavy-ion collisions observable,
such as rapidity spectra of net proton production, transverse
momentum distributions and transverse momentum fluctuations.
Moreover, we study the nuclear and the subnuclear equation of state,
investigating the critical densities of a phase transition to a
hadron-quark-gluon mixed phase by requiring the Gibbs conditions on
the global conservation of the electric and the baryon charges. The
relevance of small deviations from the standard extensive statistics
is studied in the context of intermediate and high energy heavy-ion
collisions.
} 
\maketitle
\section{Introduction}

It is common opinion that hadrons dissociate into a plasma of their
elementary constituents, quarks and gluons (QGP), at density several
times the nuclear matter density and/or at temperature above few
hundreds MeV, which is the critical temperature $T_c$ of the transition
from the QGP phase to the hadronic gas phase and viceversa. Such a QGP
is expected to have occurred in the early stages of the Universe and can
be found in dense and hot stars, neutron stars,
nucleus-nucleus high energy collisions where heavy ions are
accelerated to relativistic energies~\cite{hwa}. After
collision, a fireball is created which may realize the conditions of the QGP. The plasma
then expands, cools, freezes-out into hadrons, photons, leptons that are
detected and analyzed \cite{biro08}.

Since the interactions among quarks and gluons become weak at small
distance or high energy, we could expect that QGP is a weakly
interacting plasma, which can be described
by perturbative QCD. However, this is rigorously true only at very
high temperature ($T>T_c$) while at the order of the critical
temperature and in the hadronization phase there are strong
non-perturbative QCD effects.

In the literature, an ordinary plasma is usually characterized by
the value of the plasma parameter $\Gamma$ \cite{ichi}
\be \Gamma=\frac{\langle U\rangle}{\langle T \rangle} \; , \ee
defined as the ratio between  potential energy $\langle U\rangle$
versus kinetic energy $\langle T \rangle$. When $\Gamma\ll 1$, one
has a dilute weakly interacting gas; the Debye screening length
$\lambda_D$ is much greater than the average interparticle distance
$r_0$ and a large number of particles is contained in the Debye
sphere. Binary collisions induced by screened forces produce, in the
classical case, the standard Maxwell-Boltzmann velocity
distribution. If $\Gamma\approx 0.1\div 1$, then $\lambda_D\approx
r_0$, and it is not possible to clearly separate individual and
collective degrees of freedom: this situation refers to a weakly
interacting, non-ideal plasma. Finally, if $\Gamma\geq 1$, the
plasma is strongly interacting, Coulomb interaction and quantum
effects dominate and determine the structure of the system.

The quark-gluon plasma close to the critical temperature is a
strongly interacting system. In fact, following
Ref.\cite{albe,peshier}, the color-Coulomb coupling parameter of the
QGP is defined, in analogy with the one of the classical plasma, as
\beq \Gamma \approx  C \frac{g^2}{4\pi r_0\, T} \; , \eeq
where $C=4/3$ or 3 is the Casimir invariant for the quarks or
gluons, respectively;  for typical temperatures attained in
relativistic heavy ion collisions, $T\simeq 200$~MeV,
$\alpha_s=g^2/(4\pi)=0.2 \div 0.5$, and $r_0\simeq n^{-1/3}\simeq
0.5$~fm ($n$ being the particle density for an ideal gas of 2 quark
flavors in QGP). Consequently, one obtains $\Gamma \simeq 1.5-5$ and
the plasma can be considered to be in a non-ideal liquid
phase~\cite{peshier,thoma1}. Furthermore, during the hadronization,
non-perturbative, confining QCD effects are important, hence:
i) near the critical temperature $T_c$, the effective quark mass is
$m_q\approx T$ and $n <r>^3 \approx n \lambda_D^3 \simeq 1$; ii) the
mean field approximation of the plasma is no longer correct and
memory effects are not negligible.

In these conditions, the generated QGP does not satisfy anymore the
basic assumptions (BBGKY hierarchy) of a kinetic equation (Boltzmann
or Fokker-Planck equation) which describes a system toward the
equilibrium. Indeed, near the phase transition the interaction
range is much larger than the Debye screening length and a small
number of partons is contained in the Debye
sphere~\cite{albe,thoma1}. Therefore, the collision time is not much
smaller than the mean time between collisions and the interaction is
not local. The binary collisions approximation is not satisfied,
memory effects and long--range color interactions give rise to the
presence of non--Markovian processes in the kinetic equation, thus
affecting the thermalization process toward equilibrium as well as
the standard equilibrium distribution.

In the last years, there has been an increasing evidence that the
generalized non-extensive statistical mechanics, proposed by Tsallis
\cite{tsallis,tsamendes,gellmann} and characterized by a power-law
stationary particle distribution, can be considered as a
 basis for a theoretical framework appropriate to incorporate, at
least to some extent and without going into microscopic dynamical
description, long-range interactions, long-range microscopic
memories and/or fractal space-time constraints. A considerable
variety of physical issues show a quantitative agreement between
experimental data and theoretical analyses based on Tsallis'
thermostatistics. In particular, there is a growing interest to high
energy physics applications of non-extensive statistics
\cite{wilk1,plb2001,rafelski,bediaga,beck,biroprl05}. Several
authors outline the possibility that experimental observations in
relativistic heavy-ion collisions can reflect non-extensive
statistical mechanics effects during the early stage of the
collisions and the thermalization evolution of the system
\cite{albe,biro04,biropla08,wilk2,wilk08,lavaphysa}.

The aim of this paper is, in the light of the recent developments,
to critically review our principal results obtained in the context
of high energy heavy-ion collisions and to gain a new deeper insight
on the nuclear equation of state and hadron-quark gluon mixed phase
in the framework of non-extensive thermostatistics.

The paper is organized as follows. We start in Sec. II with a short
reminder on the relativistic non-extensive thermodynamics. In Sec.
III, we review the phenomenological studies on rapidity distribution
of the net proton production; Sec. IV and in Sec. V are devoted to
transverse observables by studying the transverse momentum
distribution and the related transverse momentum fluctuations. In
Sec. VI we investigate the effects of non-extensive thermostatistics
on hadronic and quark-gluon equation of state; afterwards, we study
the formation of hadron-quark mixed phase on the basis of the Gibbs
condition in which both baryon and isospin charge are preserved.

\section{Relativistic non-extensive thermodynamics}

In order to study from a phenomenological point of view experimental
observable in relativistic heavy-ion collisions, in this Section we
present the basic macroscopic thermodynamic variables and kinetic
theory in the language of the non-extensive relativistic kinetic
theory.

Let us start by introducing the particle four-flow in the
phase space as \cite{lavapla}
\begin{equation}
N^\mu(x)=\frac{1}{Z_q}\int \frac{d^3p}{p^0} \, p^\mu \,f(x,p) \; ,
\label{nmu}
\end{equation}
and the energy-momentum flow as
\begin{equation}
T^{\mu\nu}(x)=\frac{1}{Z_q}\int \frac{d^3p}{p^0} \, p^\mu p^\nu \,
f^q(x,p) \; , \label{tmunu}
\end{equation}
where we have set $\hbar=c=1$, $x\equiv x^\mu=(t,{\bf x})$, $p\equiv
p^\mu=(p^0,{\bf p})$, $p^0=\sqrt{{\bf p}^2+m^2}$ being the
relativistic energy and $f(x,p)$  the particle distribution
function. The four-vector $N^\mu=(n,{\bf j})$ represents the
probability density $n=n(x)$ (which is normalized to unity) and the
probability flow ${\bf j}={\bf j}(x)$. The energy-momentum tensor
contains the normalized $q$-mean expectation value of the energy
density, as well as the energy flow, the momentum and the momentum
flow per particle. Its expression follows directly from the
definition of the mean $q$-expectation value in non-extensive
statistics \cite{tsamendes}; for this reason $T^{\mu\nu}$ it is
given in terms of $f^q(x,p)$.

On the basis of the above definitions, one can show that it is
possible to obtain a generalized non-linear relativistic Boltzmann
equation \cite{lavapla}
\begin{equation}
p^\mu \partial_{\mu}\left[f(x,p)\right]^q=C_q(x,p)  \; ,
\label{boltz}
\end{equation}
where the function $C_q(x,p)$  implicitly defines a generalized
non-extensive collision term
\begin{eqnarray}
C_q(x,p)=&&\frac{1}{2}\! \int\!\!\frac{d^3p_1}{p^0_1}
\frac{d^3p{'}}{p{'}^0} \frac{d^3p{'}_1}{p{'}^0_1} \Big
\{h_q[f{'},f{'}_1]  W(p{'},p{'}_1\vert p,p_1) \nonumber \\
&&-h_q[f,f_1]  W(p,p_1\vert p{'},p{'}_1) \Big\}\; .
\end{eqnarray}
Here $W(p,p_1\vert p{'},p{'}_1)$ is the transition rate between a
two-particle state with initial four-momenta $p$ and $p_1$ and a
final state with four-momenta $p{'}$ and $p{'}_1$; $h_q[f,f_1]$ is
the $q$-correlation function relative to two particles in the same
space-time position but with different four-momenta $p$ and $p_1$,
respectively. Such a transport equation conserves the probability
normalization (number of particles) and is consistent with the
energy-momentum conservation laws in the framework of the normalized
$q$-mean expectation value. Moreover, the collision term contains a
generalized expression of the molecular chaos and for $q>0$ implies
the validity of a generalized $H$-theorem, if the following,
non-extensive, local four-density entropy is assumed (henceforward
we shall set Boltzmann constant $k_{_B}$ to unity)
%
\begin{equation}
S_q^\mu(x)=- \int \frac{d^3p}{p^0} \,p^\mu f[(x,p)]^q [\ln_q
f(x,p)-1] \; , \label{entro4}
\end{equation}
where we have used the definition $\ln_q x=(x^{1-q}-1)/(1-q)$, the
inverse function of the Tsallis' $q$-exponential function
\begin{equation}
e_q(x)=[1+(1-q)x]^{1/(1-q)} \; , \label{tsaexp}
\end{equation}
which satisfies the property $e_q(\ln_q x)=x$.

The above expression is written in a covariant form, in fact
$S^{\mu}_q=(S^0_q,S^i_q)$, with $i=1,2,3$, correctly transforms as a
four-vector under Lorentz transformations~\cite{lavapla}, where
$S^0_q$ is the standard expression of the Tsallis non-extensive
local entropy density and $S^i_q$ is the Tsallis entropy flow. Note
that for $q\rightarrow 1$, Eq.(\ref{entro4}) reduces to the well
known four-flow entropy expression \cite{groot}.

At equilibrium, the solution of the above Boltzmann equation is a
relativistic Tsallis-like (power law) distribution and can be
written as
\begin{equation}
f_{eq}(p)= \frac{1}{Z_q}\left [1-(1-q) \frac{p^\mu U_\mu}{T}
\right]^{1/(1-q)} \; , \label{nrdistri}
\end{equation}
where $U_\mu$ is the hydrodynamic four-velocity \cite{groot} and
$f_{eq}$ depends only on the momentum in the absence of an external
field. At this stage, $T$ is a free parameter and only in the
derivation of the equation of state it will be identified with the
physical temperature.

We are able now to evaluate explicitly all other thermodynamic
variables and provide a complete macroscopic description of a
relativistic system at the equilibrium. Considering the
decomposition of the energy-momentum tensor: $T^{\mu\nu}=\epsilon\,
U^\mu U^\nu-P\, \Delta^{\mu\nu}$, where $\epsilon$ is the energy
density, $P$ the pressure and $\Delta^{\mu\nu}=g^{\mu\nu}-U^\mu
U^\nu$, the equilibrium pressure can be calculated as
\begin{equation}
P=-\frac{1}{3} T^{\mu\nu}\Delta_{\mu\nu}=-\frac{1}{3\,Z_q}
\int\frac{d^3p}{p^0} p^\mu p^\nu
\Delta_{\mu\nu}\fq_{\!\!\!\!\!eq}(p) \; .  \label{pressrel}
\end{equation}
Setting $\tau=p^0/T$ and $z=m/T$, the above integral can
be expressed as
\begin{equation}
P=\frac{4\pi}{Z_q}\,m^2\,T^2\, K_2(q,z) \; ,\label{press}
\end{equation}
where we have introduced the $q$-modified Bessel function of the
second kind as follows
\begin{equation}
K_n(q,z)=\frac{2^n n!}{(2n)!}\frac{1}{z^n}\int^\infty_z \!\!d\tau
(\tau^2-z^2)^{n-1/2}\, \left(e^{-\tau}_q\right)^q \;
,\label{qbessel}
\end{equation}
and $e_q(x)$ is the $q$-modified exponential defined in
Eq.(\ref{tsaexp}).

Similarly, the energy density $\epsilon$ can be obtained from the
following expression
\begin{equation}
\epsilon=T^{\mu\nu}U_\mu U_\nu=\frac{1}{Z_q}  \int\frac{d^3p}{p^0}
(p^\mu U_\mu)^2 \fq_{\!\!\!\!\!eq}(p) \; , \label{energyrel}
\end{equation}
and, after performing the integration, it can be cast into the compact
expression:
\begin{equation}
\epsilon=\frac{4\pi}{Z_q}\, m^4 \left [
3\frac{K_2(q,z)}{z^2}+\frac{K_1(q,z)}{z}\right ] \; .
\end{equation}
Thus the energy per particle $e=\epsilon/n$ is
\begin{equation}
e=3 \,T +m \,\frac{K_1(q,z)}{K_2(q,z)} \; ,
\end{equation}
which has the same structure of the relativistic expression obtained
in the framework of the equilibrium Boltzmann-Gibbs statistics
\cite{groot}.

It is also interesting to consider the ratio $\epsilon/p$, a quantity often
considered in lattice calculations as an indicator for the phase transition~\cite{Allton:2003vx}:
\begin{equation}
\frac{\epsilon}{P}= 3\left(1+\frac{z}{3}\frac{K_1(q,z)}{K_2(q,z)}\right) \; ,
\end{equation}
which tends to the Stefan-Boltzmann limit 3 in the limit of very large temperatures.

In the non-relativistic limit ($p\ll 1$) the energy per particle
reduces to the well-known expression
\begin{equation}
e\simeq m+\frac{3}{2}\,T  \; ,
\end{equation}
and no explicit $q$-dependence is left over.

Hence from the above results
it appears that, in searching for the relevance of non-extensive statistical effects,
both microscopic observable, such as particle distribution, correlation functions,
fluctuations of thermodynamical variables, and macroscopic variables, such as energy
density or pressure, can be affected by the deformation parameter $q$.

In this context, it appears relevant to observe that, in Ref.~\cite{biroprl05}
non-extensive Boltzmann equation has been studied
and proposed for describing the hadronization of quark matter.
Moreover, starting from the above generalized relativistic kinetic
equations, in Ref.\cite{wilk08} the authors have recently formulated
a non-extensive hydrodynamic model for multiparticle production
processes in relativistic heavy-ion collisions. These works
represent an important bridge for a close connection between a
microscopic non-extensive model and experimental observable.

Finally, let us remind the reader that for a system of particles in a degenerate
regime the above classical distribution function (\ref{nrdistri})
has to be modified by including the fermion and boson quantum
statistical prescriptions. For a dilute gas of particles and/or for
small deviations from the standard extensive statistics ($q\approx
1$) the equilibrium distribution function, in the grand canonical
ensemble, can be written as \cite{buyu}
\begin{equation}
n^q(k,\mu)=\frac{1} { [1+(q-1)(E(k)-\mu)/T ]^{1/(q-1)} \pm 1}  \, ,
\label{distrifd}
\end{equation}
where the sign $+$ stands for fermions and $-$ for bosons: hence
all previous results can be easily extended to the case of
quantum statistical mechanics.

\section{Rapidity distributions}

Recent results for net-proton rapidity spectra in central Au+Au
collisions at the highest RHIC energy of $\sqrt{s_{NN}}$= 200 GeV
\cite{brahms} show an unexpectedly large rapidity density at
midrapidity in comparison with analogous spectra at lower energy at
SPS \cite{na49} and AGS \cite{ags}. As outlined from different
authors, such spectra can reflect non-equilibrium effects even if
the energy dependence of the rapidity spectra is not very well
understood \cite{brahms,giapu}.

In this Section we are going to discuss from a phenomenological
point of view the relevance of non-extensive statistical mechanics
and anomalous diffusion in a power law Fokker-Planck kinetic
equation which describes the non-equilibrium evolution of the
rapidity spectra of net proton yield. Let us recall, in this
context, that similar approaches have been considered in the past to
analyze transverse momentum distributions, power law spectra at
large $p_\perp$ and two-particle Bose-Einstein correlation functions
in terms of various non-conventional extensions of the
Boltzmann-Gibbs thermostatistics \cite{biro04,wilk2,csorgo1}.
Relevant results have been also obtained by Wolschin
\cite{wol2,wol4,wol08} within a three-component relativistic
diffusion model.

In order to study the rapidity spectra, it is convenient to separate
the kinetic variables into their transverse and longitudinal
components, the latter being related to the rapidity
$y=\tanh^{-1}\left(p_\parallel/\sqrt{m^2+p^2}\right)$.
If we assume that the particle distribution function $f(y,m_\perp,t)$, at
fixed transverse mass $m_\perp=\sqrt{m^2+p_\perp^2}$, is not
appreciably influenced by the transverse dynamics (which is
considered in thermal equilibrium), the non-linear Fokker-Planck
equation in the rapidity space $y$ can be written as
\cite{physicaA2008}
\begin{eqnarray}
\frac{\partial}{\partial
t}[f(y,m_\perp,t)]&&=\frac{\partial}{\partial
y} \Big [ J(y,m_\perp) [f(y,m_\perp,t)]+\nonumber\\
&&D \frac{\partial}{\partial y}[f(y,m_\perp,t)]^\mu \Big]
\label{nlfpe} \; ,
\end{eqnarray}
where $D$ and $J$ are the diffusion and drift coefficients,
respectively, while $\mu$ is a generic, real exponent.

Tsallis and Bukman \cite{tsabu} have shown that, for linear drift,
the time dependent solution of the above equation is a Tsallis
distribution with $\mu=2-q$ and that a value of $q\ne 1$ implies
anomalous diffusion, i.e., $[y(t)-y_M(t)]^2$ scales like $t^\alpha$,
with $\alpha=2/(3-q)$. For $q<1$, the above equation implies
anomalous sub-diffusion, while for $q>1$, we have a super-diffusion
process in the rapidity space. Let us observe that, at variance with
our approach, if one assumes a Fokker-Planck equation with
fractional derivatives, in the framework of the so-called continuous
time random (L\'evy) walk models, anomalous diffusion processes can
be also realized \cite{klafter,fract2,csorgo2}.

Let us observe that the choice of the diffusion and the drift
coefficients plays a crucial r\^ole in the solution of the above
non-linear Fokker-Planck equation (\ref{nlfpe}). Such a choice
influences the time evolution of the system and its equilibrium
distribution. By imposing the validity of the Einstein relation for
Brownian particles, we can generalize to the relativistic case the
standard expressions of diffusion and drift coefficients as follows
\begin{equation}
D=\gamma \, T\; , \ \ \ \  \ J(y,m_\perp)=\gamma \, m_\perp \sinh(y)
\equiv  \gamma \, p_\parallel \; , \label{coeffi}
\end{equation}
where $p_\parallel$ is the longitudinal momentum, $T$ is the
temperature and $\gamma$ is a common constant. Let us remark that
the above definition of the diffusion and drift
 coefficients appears as the natural generalization to the relativistic Brownian case in
the rapidity space. The drift coefficient which is linear in the
longitudinal momentum $p_\parallel$ becomes non-linear in the
rapidity coordinate.

It is easy to see that the above coefficients give us the Boltzmann
stationary distribution in the linear case ($q=\mu=1$), while the
equilibrium solution $f^{eq}(y,m_\perp)$ of Eq.(\ref{nlfpe}), with
$\mu=2-q$, is a Tsallis-like (power-law) distribution with the
relativistic energy $E=m_\perp \cosh(y)$
\begin{equation}
f^{eq}(y,m_\perp)\propto \Big [1-(1-q)\, m_\perp \cosh(y)/T \Big
]^{1/(1-q)} \; .
\end{equation}

The rapidity distribution at fixed time can be obtained out of
equilibrium by means of numerical integration of Eq.(\ref{nlfpe})
with delta function initial conditions depending upon the value of
the experimental projectile rapidities and by means of numerical
integration over the transverse mass $m_\perp$
\begin{equation}
\frac{dN}{dy}(y,t)=c\, \int_m^\infty \!\! m^2_\perp \, \cosh(y) \,
f(y,m_\perp,t) \, dm_\perp \; , \label{raint}
\end{equation}
where $m$ is the mass of the considered particles and $c$ is a
normalization constant, fixed by comparison with the experimental
data. The calculated rapidity spectra will ultimately depend on two
parameters: the ``interaction'' time $\tau_{int}=m\,\gamma \,t$ and
the non-extensive parameter $q$.

It is important to note that, as we will see in the next Section by
studying the transverse mass spectrum, dynamical collective
interactions are intrinsically involved in the generalized
non-extensive statistical mechanics and, in a purely thermal source,
a generalized $q$-blue shift factor (strictly related to the
presence of longitudinal flow) appears. In this context, it is worth mentioning
that collective transverse flow effects in the framework of a
non-extensive statistical mechanics have been investigated in
Ref.~\cite{biro04} as well.

\begin{figure}
\resizebox{0.45\textwidth}{!}{%
  \includegraphics{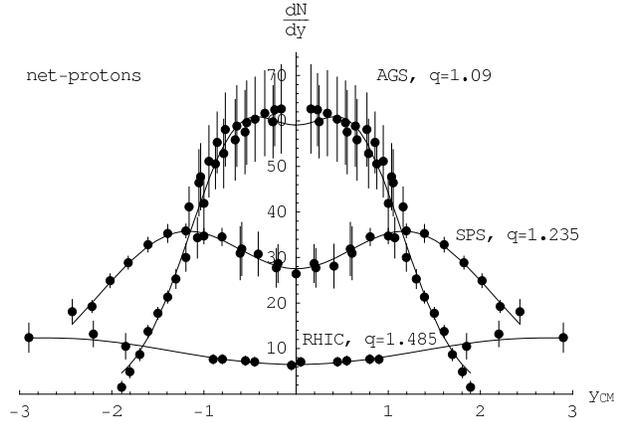}
} \caption{Rapidity spectra for net proton production
($p-\overline{p}$) at RHIC (Au+Au at $\sqrt{s_{NN}}=200$ GeV, BRAHMS
data), SPS (Pb+Pb at $\sqrt{s_{NN}}=17.3$ GeV, NA49 data) and AGS
(Au+Au at $\sqrt{s_{NN}}=5$ GeV, E802, E877, E917).}
\label{fig_rapitot}
\end{figure}

In Fig.~\ref{fig_rapitot}, we report the rapidity
distribution obtained from Eq.~(\ref{raint}) (full line) for the net proton production
($p-\overline{p}$) compared with the experimental data of RHIC
(Au+Au at $\sqrt{s_{NN}}=200$ GeV, \cite{brahms}), SPS (Pb+Pb at
$\sqrt{s_{NN}}=17.3$ GeV, \cite{na49}) and AGS (Au+Au at
$\sqrt{s_{NN}}=5$ GeV, \cite{ags}).
 The parameters employed for the three curves are: $q=1.485$ with
$\tau_{int}=0.47$ for RHIC, $q=1.235$ with $\tau_{int}=0.84$ for SPS
and $q=1.09$ with $\tau_{int}=0.95$ for AGS, respectively. Let us
notice that the value $\tau_{int}=0.47$ is compatible with
the equilibration time extracted from a hydro-description of the
RHIC data: this partly justifies the present use of near-equilibrium
distributions.

\begin{figure}
\resizebox{0.45\textwidth}{!}{%
  \includegraphics{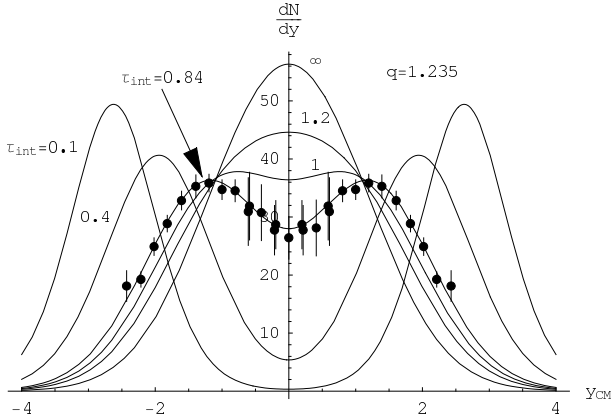}
} \caption{Time evolution of the rapidity spectra at fixed value of
$q=1.235$, obtained for SPS data. The data can be well reproduced
only for this value of $q$ and for the corresponding value of
$\tau_{int}=0.84$.} \label{fig_rapitime}
\end{figure}

We also remark that, although $q$ and $\tau_{int}$ appear, in
principle, as independent parameters, in fitting the data they are
not. We can see that in the non-linear case only ($q\ne 1$)
there exists indeed one (and only one) finite time  $\tau_{int}$ for which
the obtained rapidity spectrum well reproduces the broad
experimental shape. On the contrary, for $q=1$, no value of
$\tau_{int}$ can be found, which allows to reproduce the data.
Evidence for this feature is shown in Fig. \ref{fig_rapitime}, where
the time evolution of the rapidity spectra is reported at the fixed
value $q=1.235$, the one employed for the SPS experiment, and
different values of $\tau_{int}$: only for $\tau_{int}=0.84$  a good
agreement with the experimental data is obtained. Moreover, for
different values of $q$ we do not find a corresponding value of the
$\tau_{int}$ parameter which allows us to reproduce the selected data.

We obtain a remarkable agreement with the experimental data by
increasing the value of the non-linear deformation parameter $q$ as
the beam energy increases. At AGS energy, the non-extensive
statistical effects are negligible and the spectrum is well
reproduced within the standard quasi-equilibrium linear approach. At
SPS energy, non-equilibrium effects and non-linear evolution become
remarkable ($q=1.235$) and such effects are even more evident for
the very broad RHIC spectra ($q=1.485$).

\begin{figure}
\resizebox{0.45\textwidth}{!}{%
  \includegraphics{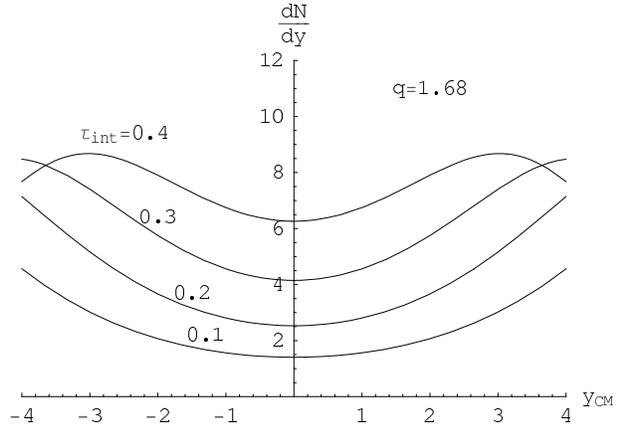}
} \caption{Rapidity spectra for net proton production expected at
LHC for the extrapolated value of $q=1.68$ and different interaction
times $\tau_{int}$. } \label{fig_rapilhc}
\end{figure}

From a phenomenological point of view, we can interpret the larger
value of the parameter $q$ and the shorter $\tau_{int}$ needed to
fit the RHIC data, as a signal of  non-linear anomalous (super)
diffusion. As confirmed by recent microscopic
calculations~\cite{peshier,thoma1}, strongly coupled non-ideal
plasma is generated at energy densities corresponding to the order
of the critical phase transition temperature and in such a regime
we find, in our macroscopic approach, strong deviations from the
standard thermostatistics. At much higher energy, such as the LHC
(Large Hadron Collider - CERN Laboratory) one, we can foresee a
minor relevance of such non-ideal effects since the considerable
expected energy density is far above the critical one.
Nevertheless we can guess, on the basis of a linear extrapolation
of the $q$-value versus the beam rapidity, that a suitable
$q$-value for LHC will be $q=1.68$. Accordingly we show in
Fig.~\ref{fig_rapilhc} the expected net-proton distributions,
evaluated at different $\tau_{int}\le 0.4$.

\section{Transverse momentum distributions}

High transverse momentum particle production in hadronic collisions
results from the fragmentation of quarks and gluons emerging from
the initial scattering at large $Q^2$, therefore, hard processes in
nucleus-nucleus collisions provide direct information on the early
partonic phases of the reaction and particle production at high
transverse momentum is sensitive to properties of the hot and dense
matter in the nuclear collisions. For this reason and for the
motivations reported in the Introduction, we expect that the
transverse momentum spectra will be sensibly affected by
non-extensive statistical effects.

The single particle spectrum can be expressed as an integral over a
freeze--out hypersurface $\Sigma_f$
\begin{equation}
E \frac{d^3N}{d^3p}=\frac{dN}{dy\, m_\perp dm_\perp d\phi}=
\frac{g}{(2\pi)^3} \int_{\Sigma_f} p^\mu d\sigma_\mu(x) f(x,p) \, ,
\end{equation}
where $g$ is the degeneracy factor and $f(x,p)$ is the phase--space
distribution.

The transverse momentum distribution depends on the phase-space
distribution and usually an exponential shape is employed to fit the
experimental data. This shape is obtained by assuming a purely
thermal source with a Boltzmann distribution and the transverse
momentum spectrum can be expressed as
\begin{equation}
\frac{dN}{m_\perp dm_\perp}=A \; m_\perp K_1\left (z \right ) \;,
\end{equation}
where $z=m_{\perp}/T$, $m_\perp=\sqrt{p_\perp^2+m^2}$, and $K_1$ is
the first order modified Bessel function. In the asymptotic limit,
$m_{\perp}\gg T$ ($z\gg 1$), the above expression gives rise to the
exponential shape
\begin{equation}
\frac{dN}{m_\perp dm_\perp}=B \; \sqrt{m_\perp} \, e^{-z} \;.
\label{mps}
\end{equation}
High energy deviations from the exponential shape can be taken into
account by introducing a dynamical effect due to collective
transverse flow, also called blue-shift, with an increase of the
slope parameter $T$ at large $m_\perp$.

Let us consider a different  point of view and argue that the
deviation from the Boltzmann slope at high $p_\perp$ can be ascribed
to the presence of non-extensive statistical effects in the steady
state distribution of the particle gas. In this framework, at the
first order in $(q-1)$ the transverse mass spectrum can be written
as \cite{albe}
\begin{eqnarray}
\frac{dN}{m_\perp dm_\perp}=&&C \; m_\perp \Big\{ K_1\left ( z
\right
)\nonumber\\
&&+ \frac{(q-1)}{8} z^2 \; [3 \; K_1 (z)+K_3 (z) ] \Big\} \;,
\label{mpt}
\end{eqnarray}
where $K_3$ is the modified Bessel function of the third order. In
the asymptotic limit, $z\gg 1$, we have
\begin{equation}
\frac{dN}{m_\perp dm_\perp}=D \; \sqrt{m_\perp} \, \exp \left
(-z+\frac{q-1}{2}\, z^2\right )\; , \label{mpst}
\end{equation}
and we may obtain the generalized slope parameter or $q$-blue shift
(if $q>1$)
\begin{equation}
T_q=T+ (q-1) \, m_\perp \; . \label{qtslope}
\end{equation}
Let us notice that the slope parameter depends on the detected
particle mass and it increases with the energy (if $q>1$) as it was
observed in the experimental results \cite{na44}.

\begin{figure}
\resizebox{0.48\textwidth}{!}{%
  \includegraphics{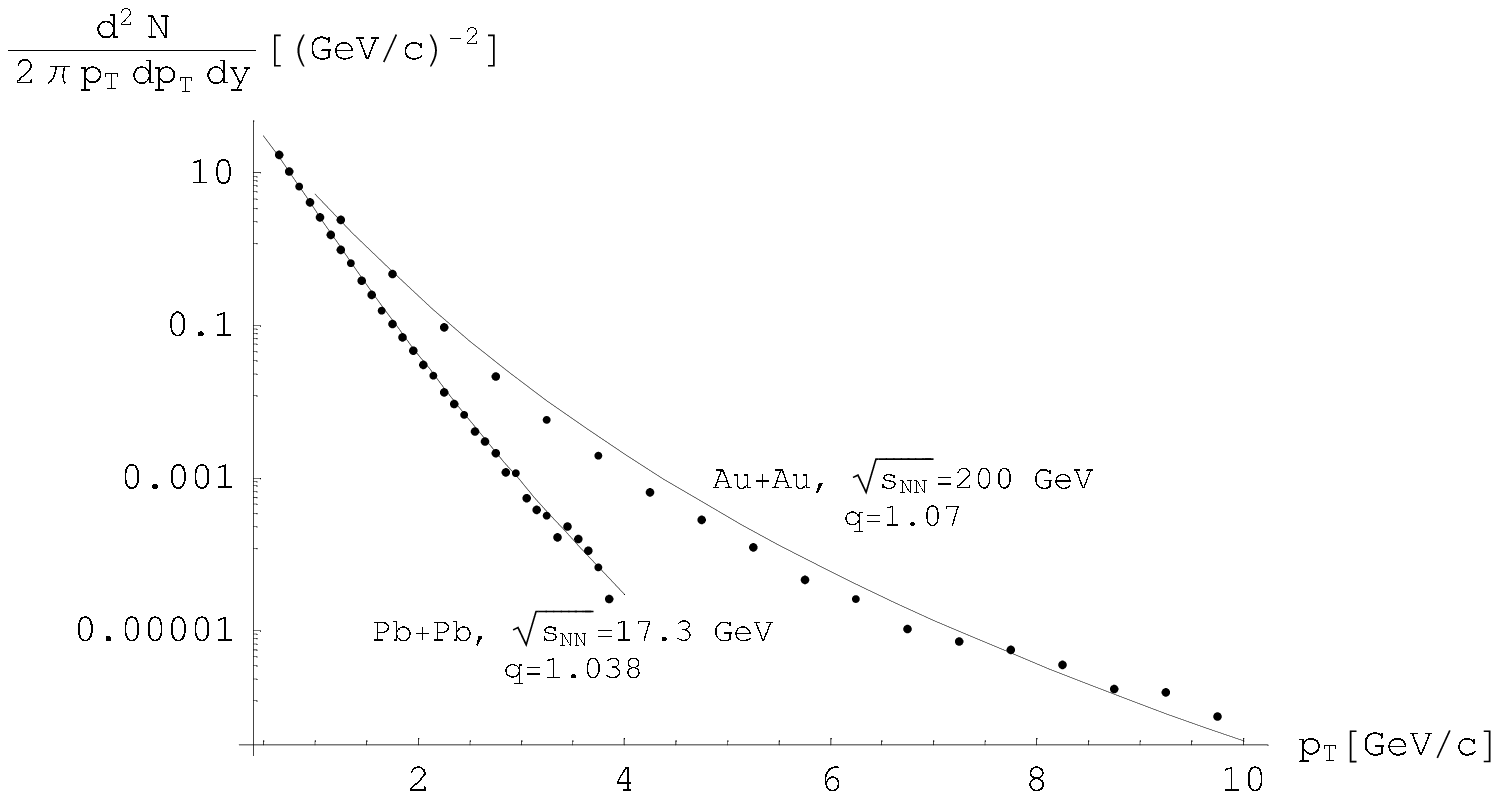}
} \caption{Experimental neutral pion invariant yields in central
Pb+Pb collisions at $\sqrt{s_{NN}}$=17.3 GeV \cite{wa98} and in
central Au+Au collisions at $\sqrt{s_{NN}}$=200 GeV \cite{phenix}
compared with the modified thermal distribution shape by using
non-extensive statistics ($q=1.038$ for Pb+Pb and $q=1.07$ for Au+Au
collisions.) } \label{figphenix}
\end{figure}

In the same manner, the particle invariant yield at midrapidity, for
central collisions and in the framework of non-extensive statistics,
can be written in terms of the equilibrium Tsallis-like distribution
(\ref{nrdistri}) as
\begin{eqnarray}
\frac{d^2N}{2 \pi p_\perp dp_\perp dy}=C \; m_\perp  \left[1-(1-q)
\frac{m_\perp}{T}\right]^{1/(1-q)} \, , \label{yields}
\end{eqnarray}
where C is a normalization constant.\\
In Fig. \ref{figphenix}, we report the experimental neutral pion
invariant yields in central Pb+Pb collisions at $\sqrt{s_{NN}}$=17.3
GeV (SPS) \cite{wa98} and in central Au+Au collisions at
$\sqrt{s_{NN}}$=200 GeV (RHIC) \cite{phenix} compared with the
modified non-exten\-sive thermal distribution shape of
Eq.(\ref{yields}). For Pb+Pb collisions we have set the
non-extensive parameter $q=1.038$ with $T$=140 MeV and for Au+Au
collisions $q=1.07$ with $T$=160 MeV. It is important to outline
that, for consistency, the same value extracted in the central Pb+Pb
collisions will be used in the next Section in the evaluation of the
transverse momentum fluctuations at the same SPS energy. Similar
results have been obtained by reproducing the experimental $S+S$
transverse momentum distribution (NA35 data \cite{na35}) in Ref.
\cite{albe} and, always in the framework of non--extensive
statistics, in Ref.\cite{wilk1}.


Let us observe that the values of the entropic $q$-parame\-ter, used
in the fit of the transverse momentum distributions, are sensibly
smaller than the ones extracted from the rapidity spectra of the
previous Section. The reason of this remarkable difference lies in
the completely different nature of the transverse and longitudinal
observables. It is rather common opinion that the longitudinal
rapidity distribution is not appreciably influenced by the
transverse dynamics, which is considered in thermal equilibrium. The
other way round, rapidity spectra is affected by non-equilibrium
features and the dynamics of the interactions, strongly related to
the value of the $q$-parameter in our macroscopic phenomenological
approach, results sensibly different.

\section{Transverse momentum fluctuations}

Ga\'zdzicki and Mr\'owczy\'nski introduced the following quantity
\cite{gaz92,mro}
\begin{equation}
\Phi_{p_\perp}=\sqrt{\langle Z_{p_\perp}^2 \rangle \over \langle N
\rangle} - \sqrt{\overline{z_{p_\perp}^2}} \;, \label{phix}
\end{equation}
where $z_{p_\perp}= p_\perp - \overline{p}_\perp$ and $Z_{p_\perp} =
\sum_{i=1}^{N}(p_{\perp i} - \overline{p}_{\perp i})$, $N$ is the
multiplicity of particles produced in a single event. Non-vanishing
$\Phi$ implies effective correlations among particles which alter
the momentum distribution.

In the framework of non--extensive statistics and keeping in mind
that it preserves the whole mathematical structure of the
thermodynamical relations,
 it is easy to show that the
two terms in the right hand side of Eq.(\ref{phix}) can be expressed
in the following simple form
\begin{eqnarray}
\overline{z^2_{p_{\perp}}} = {1 \over \rho}\int{d^3p \over (2\pi
)^3} \, \, \Big(p_{\perp} - \overline{p}_{\perp} \Big)^2   \langle n
\rangle_q \; , \label{qz2p}
\end{eqnarray}
and
\begin{eqnarray}
{\langle Z_{p_{\perp}}^2 \rangle \over \langle N \rangle }= {1 \over
\rho}\int{d^3p \over (2\pi )^3} \,\Big(p_{\perp} -
\overline{p}_{\perp} \Big)^2 \langle\Delta n^2\rangle_q \;,
\label{qz2pn}
\end{eqnarray}
where
\begin{equation}
\overline{p}_{\perp} = {1 \over \rho}\int{d^3p \over (2\pi )^3} \;
p_{\perp} \langle n \rangle_q
\end{equation}
with
\begin{equation}
\rho=\int{d^3p \over (2\pi )^3} \langle n \rangle_q \;.
\end{equation}

In the above equations we have indicated with $\langle n \rangle_q$
the following mean occupation number of bosons \cite{buyu}
\begin{equation}
\langle n\rangle_q=\frac{1}{[1+(q-1)\beta (E-\mu)]^{1/(q-1)}- 1} \;,
\label{distri}
\end{equation}
and with $\langle\Delta n^2\rangle_q=\langle n^2\rangle_q-\langle n
\rangle^2_q$ the generalized particle fluctuations, given by
\begin{eqnarray}
\langle\Delta n^2\rangle_q\equiv
\frac{1}{\beta}\frac{\partial\langle n\rangle_q}{\partial\mu}&=&
\frac{\langle n\rangle_q }{1+(q-1)\beta (E-\mu)}\, (1\mp \langle
n\rangle_q)\nonumber\\
&=&\langle n\rangle_q^q \; (1 \mp \langle n\rangle_q )^{2-q}.
\label{fluc}
\end{eqnarray}

NA49 Collaboration has measured the correlation $\Phi_{p_\perp}$ of
the pion transverse momentum (Pb+Pb at 158 A GeV) \cite{na49fluc}
obtaining $\Phi^{exp}_{p_\perp}=(0.6\pm 1) \; {\rm MeV}$. This value
is the sum of two contributions: $\Phi^{st}_{p_\perp}=(5\pm 1.5) \;
{\rm MeV}$, the measure of the statistical two-particle correlation,
and $\Phi^{tt}_{p_\perp}=(-4\pm 0.5) \; {\rm MeV}$, the
anti-correlation from limitation in two-track resolution.

Standard statistical calculations ($q=1$) give \cite{mro}
$\Phi^{st}_{p_\perp} = 24.7\; {\rm MeV \; at \; } T = 170 \; {\rm
MeV}, \ \ \mu = 60 \; {\rm MeV}$. In the frame of non--extensive
statistics, for $q=1.038$ (for consistency, the same value extracted
in the previous Section from neutral pion invariant yield for
central Pb+Pb collisions at the same energy in consideration), we
obtain the experimental (statistical) value: $\Phi^{st}_{p_\perp} =
5 \; {\rm MeV \; at\;\,} T = 170 \; {\rm MeV}$ and $\mu = 60 \; {\rm
MeV}$.

\begin{figure}
\resizebox{0.45\textwidth}{!}{%
  \includegraphics*[70,240][550,560]{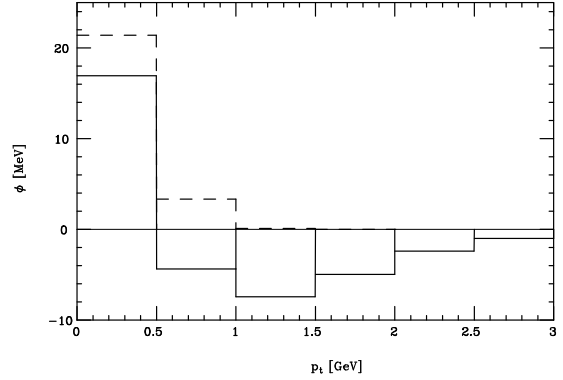}
} \caption{The partial contributions to the correlation measure
$\Phi_{p_\perp}$ [MeV] in different $p_\perp$ intervals. The dashed
line refers to standard statistical calculations with $q=1$, the
solid line corresponds to $q=1.038$. } \label{fig_isto}
\end{figure}

In Fig. \ref{fig_isto}, we show the partial contributions to the
quantity $\Phi_{p_\perp}$, by using Eq.s (\ref{qz2p}) and
(\ref{qz2pn}), and by extending the integration over $p_\perp$ to
partial intervals $\Delta p_\perp=0.5$~GeV at $T=170$ MeV and
$\mu=60$ MeV. In the standard statistics (dashed line),
$\Phi_{p_\perp}$ is always positive and vanishes in the
$p_\perp$-intervals above $\approx 1$ GeV. In the non-extensive
statistics (solid line), instead, the fluctuation measure
$\Phi_{p_\perp}$ becomes negative for $p_\perp$ larger than $0.5$
GeV and becomes vanishingly small only in $p_\perp$-intervals
above $\sim 3$ GeV. If measured in separate $p_\perp$ bins, such a
negative value of $\Phi_{p_\perp}$ at high $p_\perp$ could be an
evidence of the presence of non-extensive regime in heavy-ions
collisions.

Finally, in this context it is important to outline that a
critical overview of the $\Phi_{p_\perp}$ measure of fluctuations
and correlations was given in Ref.s
\cite{wilk-prc2001,wilk-acta2004}. It was shown that
$\Phi_{p_\perp}$ measure is very sensitive to the constraints
provided by the energy-momentum conservation laws and to the
effects of correlations. These effects should be carefully
accounted in the phenomenological studies related to event by
event analysis of data.

\section{Non-extensive nuclear equation of state}

As partially discussed in the Introduction, hadronic matter is
expected to undergo a phase transition into a deconfined phase of
quarks and gluons at large densities and/or high temperatures.
However, the extraction of experimental information about the
Equation of State (EOS) of matter at large densities and
temperatures from the data of intermediate and high energy heavy-ion
collisions is very complicated. Possible indirect indications of a
softening of the EOS at the energies reached at AGS have been
discussed several times in the literature \cite{stoecker,prl}. In
particular, a recent analysis \cite{ivanov} based on a 3-fluid
dynamics simulation suggests a progressive softening of the EOS
tested through heavy-ion collisions at energies ranging from 2A GeV
up to 8A GeV. On the other hand, the information coming from
experiments with heavy-ions at intermediate and high energy
collisions is that, for symmetric or nearly symmetric nuclear
matter, the critical density (at low temperatures) appears to be
considerably larger than nuclear matter saturation density $\rho_0$.
Concerning non-symmetric matter, general arguments based on Pauli
principle suggest that the critical density decreases with $Z/A$.
Therefore, the transition's critical densities are expected to
sensibly depend on the isospin of the system \cite{ditoro}.
Moreover, the analysis of observations of neutron stars, which are
composed of $\beta$-stable matter for which $Z/A\le 0.1$, can also
provide hints on the structure of extremely asymmetric matter at
high density. No data on the quark deconfinement transition are at
the moment available for intermediate values of $Z/A$. Recently, it
has been proposed by several groups to produce unstable neutron-rich
beams at intermediate energies. These new experiments can open the
possibility to explore in laboratory the isospin dependence of the
critical densities.

The aim of this Section is to study the behavior of the nuclear
equation of state at finite temperature and baryon density and to
explore the existence of a hadron-quark mixed phase at a fixed value
of $Z/A$. Furthermore, from the above considerations, it appears
reasonable that in regime of high density and temperature both
hadron and quark-gluon EOS can be sensibly affected by non-extensive
statistical effects \cite{physicaEOS}. The relevance of
these effects on the relativistic hadronic
equation of state has also been recently point out in Ref.~\cite{silva}.

The scenario we are going to explore in this last Section
corresponds to the situation realized in experiments at not too high
energy. In this condition, only a small fraction of strangeness can
be produced and, therefore, we limit ourselves to study the
deconfinement transition from nucleonic matter into up and down
quark matter. In the next two subsections, we will study the two
corresponding EOSs separately, on the basis on the previously
reported non-extensive relativistic thermodynamic relations. The
existence of the hadron-quark mixed phase will be studied in the
third subsection. This investigation may be helpful also in view of
the future experiments planned, e.g., at the facility FAIR at GSI
\cite{senger}.

\subsection{Non-extensive hadronic equation of state}

Concerning the hadronic phase, we use a relativistic self-consistent
theory of nuclear matter in which nucleons interact through the
nuclear force mediated by the exchange of virtual isoscalar and
isovector mesons ($\sigma,\omega,\rho$) \cite{glen}. On the basis of
the Eqs.(\ref{tmunu}), (\ref{pressrel}) and (\ref{energyrel}), the
pressure and the energy density can be written as
\begin{eqnarray}
P&=&\sum_{i=n,p} \frac{2}{3}\int \frac{{\rm d}^3k}{(2\pi)^3}
\frac{k^2}{E_{i}^\star(k)}
[n_i^q(k,\mu_i^\star)+n_i^q(k,-\mu_i^\star)]
 \nonumber \\
&-&\frac{1}{2}m_\sigma^2\sigma^2 - \frac{1}{3} a
\sigma^3-\frac{1}{4}b\sigma^4+
\frac{1}{2}m_\omega^2\omega_0^2+\frac{1}{2}m_{\rho}^2 \rho_0^2,\\
\epsilon&=& \sum_{i=n,p}{2}\int \frac{{\rm
d}^3k}{(2\pi)^3}E_{i}^\star(k)
[n_i^q(k,\mu_i^\star)+n_i^q(k,-\mu_i^\star)]\nonumber \\
&+&\frac{1}{2}m_\sigma^2\sigma^2+\frac{1}{3} a
\sigma^3+\frac{1}{4}b\sigma^4
+\frac{1}{2}m_\omega^2\omega_0^2+\frac{1}{2}m_{\rho}^2 \rho_0^2,
\end{eqnarray}
where $n_i(k,\mu_i)$ and $n_i(k,-\mu_i)$ are the fermion particle
and antiparticle distribution (\ref{distrifd}). The nucleon
effective energy is defined as
${E_i}^\star=\sqrt{k^2+{{M_i}^\star}^2}$, where
${M_i}^\star=M_{i}-g_\sigma \sigma$. The effective chemical
potentials $\mu_i^\star$  are given in terms of the vector meson
mean fields $\mu_i^\star=\mu_i-g_\omega\omega_0\mp g_{\rho}\rho_0$
($-$ proton, $+$ neutron), where $\mu_i$ are the thermodynamical
chemical potentials $\mu_i=\partial\epsilon/\partial\rho_i$. At
zero temperature they reduce to the Fermi energies $\mu_i^\star =
E_{Fi}^\star \equiv \sqrt{k_{Fi}^2+{M_i^\star}^2}$ and the
non-extensive statistical effects disappear. The isoscalar and
isovector meson fields ($\sigma$, $\omega$ and $\rho$) are
obtained as a solution of the field equations in mean field
approximation and the related couplings ($g_\sigma$, $g_\omega$
and $g_\rho$) are the free parameter of the model \cite{glen}.
Finally, The baryon density $\rho_B$ is given by
\begin{equation}
\rho_B=2 \sum_{i=n,p} \int\frac{{\rm
d}^3k}{(2\pi)^3}[n_i(k,\mu_i^\star)-n_i(k,-\mu_i^\star)]\,.
\end{equation}

Note that statistical mechanics enters as an external ingredient in
the functional form of the "free" particle distribution of Eq.~(\ref{distrifd}).
Since all the equations must be solved in a
self-consistent way, the presence of non-extensive statistical
effects in the particle distribution function influences the
many-body interaction in the mean field self-consistent solutions
obtained for the meson fields.

\begin{figure}
\resizebox{0.55\textwidth}{!}{%
\includegraphics*[90,550][500,800]{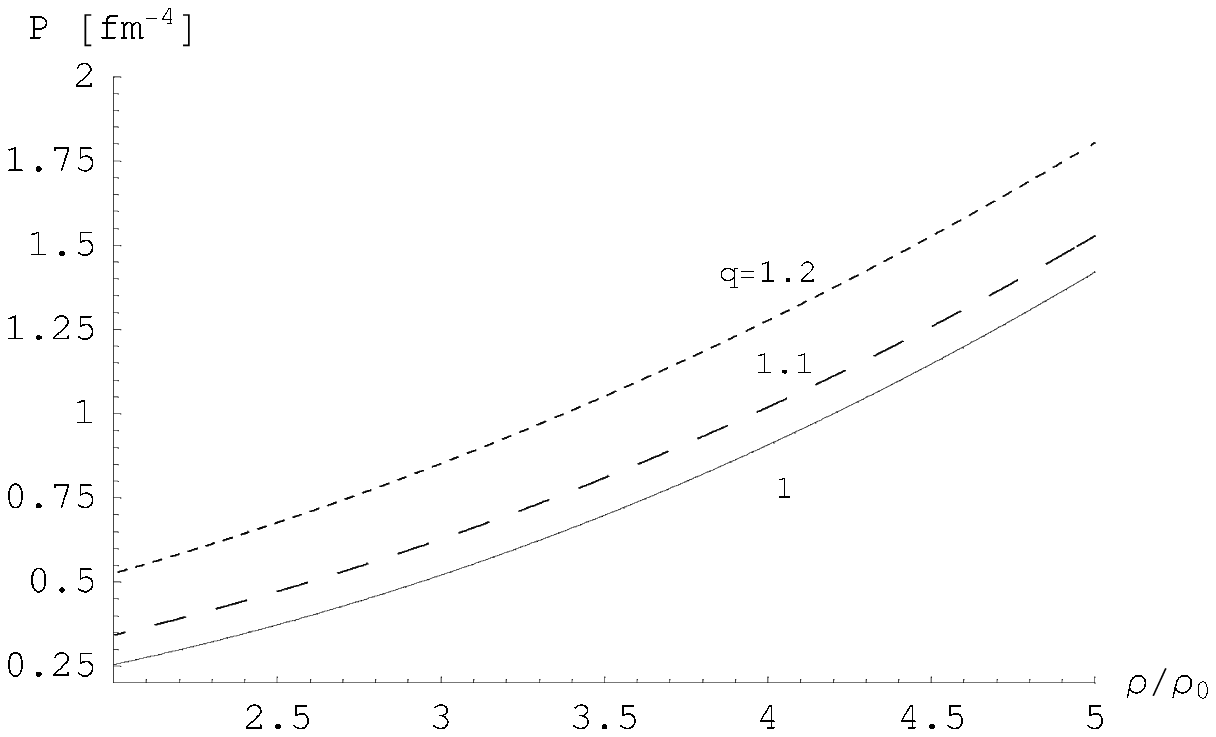}
} \caption{Hadronic equation of state: pressure versus baryon number
density (in units of the nuclear saturation density $\rho_0$) for
different values of $q$. In the figure $T=100$ MeV and $Z/A=0.4$.} \label{figph}
\end{figure}

In Fig. \ref{figph}, we report the resulting hadronic EOS: pressure
as a function of the baryon number density for different values of
$q$. Since in the previous Sections we have phenomenologically
obtained values of $q$ greater than unity, we will concentrate our
analysis to $q>1$. The results are plotted at the temperature
$T=100$ MeV, at fixed value of $Z/A=0.4$ and we have used the GM2
set of parameters of Ref.\cite{glen}. The range of the considered
baryon density and the chosen values of the parameters correspond to
a physical situation which can be realized in the recently proposed high
energy heavy-ion collisions experiment at GSI~\cite{gsi}.

\subsection{Non-extensive QGP equation of state}

In the simple model of free quarks in a bag \cite{mit}, the
pressure, energy density and baryon number density for a relativistic Fermi gas
of quarks in the framework of non-extensive statistics (see
Eqs.(\ref{nmu}), (\ref{tmunu}), (\ref{pressrel}) and
(\ref{energyrel})) can be written, respectively, as
\begin{eqnarray}
&P& =\sum_{f=u,d} \frac{1}{3}\frac{\gamma_f}{2\pi^2} \int^\infty_0
k \frac{\partial\epsilon_f}{\partial k}
[n_f^q(k,\mu_f)+n_f^q(k,-\mu_f)]
k^2 dk \nonumber \\&&\;\;\;\;-B, \label{bag-pressure}\\
&\epsilon& =\sum_{f=u,d} \frac{\gamma_f}{2\pi^2} \int^\infty_0
\epsilon_f
[n_f^q(k,\mu_f)+n_f^q(k,-\mu_f)] k^2 dk\label{bag-energy} \nonumber \\&&\;\;\;\;+B, \\
&\rho& =\sum_{f=u,d} \frac{1}{3}\frac{\gamma_f}{2\pi^2}
\int^\infty_0 [n_f(k,\mu_f)-n_f(k,-\mu_f)] k^2 dk\, ,
\label{bag-density}
\end{eqnarray}
where $\epsilon_f=(k^2+m_f^2)^{1/2}$ and $n_f(k,\mu_f)$,
$n_f(k,-\mu_f)$ are the particle and antiparticle quark
distributions. The quark degeneracy for each flavor is
$\gamma_f=6$. Similar expressions for the pressure and the energy
density can be written for the gluons treating them as a massless
Bose gas with zero chemical potential and degeneracy factor
$\gamma_g=16$. In this subsection we are limiting our study to the
two-flavor case ($f=u, \, d$). As already remarked, this appears
rather well justified for the application to heavy ion collisions
at relativistic (but not ultra-relativistic) energies, the
fraction of strangeness produced at these energies being small
\cite{ditoro2,fuchs}.

Since one has to employ the fermion (boson) non-exten\-sive distribution
(\ref{distrifd}), the results are not analytical, even in the
massless quark approximation. Hence a numerical evaluations of
the integrals in Eq.s~(\ref{bag-pressure})--(\ref{bag-density}) must be performed.
A similar calculation,
only for the quark-gluon phase, was also performed in
Ref.\cite{miller} by studying the phase transition diagram.

\begin{figure}
\resizebox{0.55\textwidth}{!}{%
  \includegraphics*[90,550][500,800]{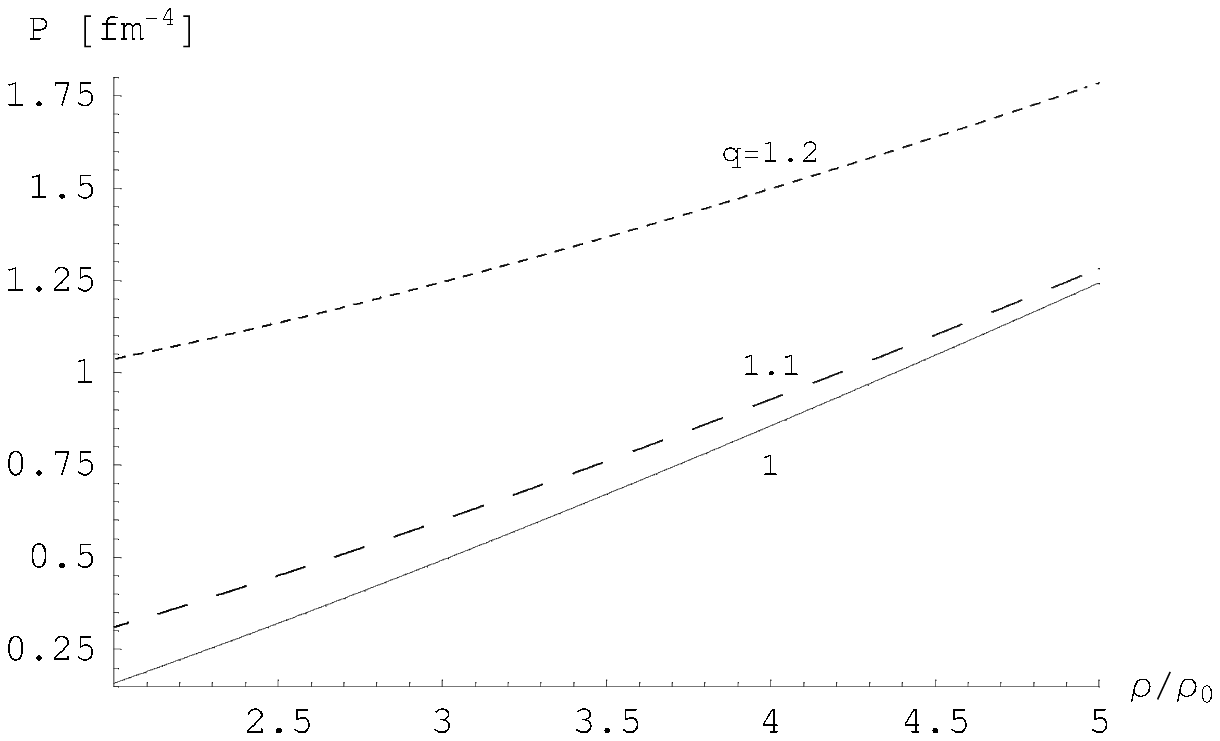}
} \caption{The same as in Fig.~\ref{figph} for the case of the quark-gluon equation
of state.} \label{figpq}
\end{figure}

In Fig. \ref{figpq}, we report the EOS for massless quarks $u$, $d$
and gluons, for different values of $q$. As in Fig.~\ref{figph}, the
results are plotted at the temperature $T=100$ MeV and at a fixed
value of $Z/A=0.4$; the bag parameter is $B^{1/4}$=170 MeV. In both
figures \ref{figph} and \ref{figpq} one can observe sizable
effects in the behaviour of the EOS even for small deviations from the
standard statistics (the largest value of $q$ employed here is 1.2).

\subsection{Mixed hadron-quark phase}

In this subsection we investigate the hadron-quark phase transition
at finite temperature and baryon chemical potential by means of the
previous relativistic EOSs. Lattice calculations predict a critical
phase transition temperature $T_c$ of about 170 MeV, corresponding
to a critical energy density $\epsilon_c\approx$ 1 GeV/fm$^3$
\cite{hwa}. In a theory with only gluons and no quarks, the
transition turns out to be of first order. In nature, since the $u$
and $d$ quarks have a small mass, while the strange quark has a
somewhat larger mass, the phase transition is predicted to be a
smooth cross over. However, since it occurs over a very narrow range
of temperatures, the transition, for several practical purposes, can
still be considered of first order. Indeed the lattice data with 2
or 3 dynamical flavours are not precise enough to unambigously
control the difference between the two situations. Thus, by
considering the deconfinement transition at finite density as a the
first order one, a mixed phase can be formed, which is typically
described using the two separate equations of state, one for the
hadronic and one for the quark phase.

To describe the mixed phase we use the Gibbs formalism, which in
Ref.~\cite{glenprd}  has been applied to systems where more than one
conserved charge is present. In this contribution we are studying
the formation of a mixed phase in which both baryon number and isospin
charge are preserved. The main result of this formalism is that, at
variance with the so-called Maxwell construction, the pressure in the mixed phase is
not constant and therefore the nuclear
incompressibility does not vanish. It is important to notice that from
the viewpoint of Ehrenfest's definition, a phase transition with two
conserved charges is considered, in the literature, not of first, but of
second order \cite{muller}.

The structure of the mixed phase is obtained by imposing the Gibbs
conditions for chemical potentials and pressure and by requiring the
global conservation of the total baryon (B ) and isospin densities
(I) in the hadronic phase (H) and in the quark phase (Q)
\begin{eqnarray}
&&\mu_B^{(H)} = \mu_B^{(Q)}\, ,\nonumber \\
&&\mu_I^{(H)} = \mu_I^{(Q)}\, ,
\nonumber \\
&& P^{(H)} (T,\mu_{B,I}^{(H)}) =  P^{(Q)} (T,\mu_{B,I}^{(Q)})\, ,
\nonumber \\
&&\rho_B=(1-\chi)\rho_B^H+\chi \rho_B^Q\,, \nonumber \\
&&\rho_I=(1-\chi)\rho_I^H+\chi \rho_I^Q\,.
\end{eqnarray}
where $\chi$ is the fraction of quark matter in the mixed phase. In
this way we can obtain the binodal surface which gives the phase
coexistence region in the $(T,\rho_B,\rho_I)$ space. For a fixed
value of the conserved charge $\rho_I$, related to the proton
fraction $Z/A \equiv (1+\rho_I/\rho_B)/2$, we study the boundaries
of the mixed phase region in the $(T,\rho_B)$ plane. We are
particularly interested in the lower baryon density border, i.e. the
critical/transition density $\rho_{cr}$, in order to check the
possibility of reaching such $(T,\rho_{cr},\rho_I)$ conditions in a
transient state during a heavy-ion collision at relativistic
energies.

\begin{figure}
\resizebox{0.55\textwidth}{!}{%
  \includegraphics*[90,550][500,800]{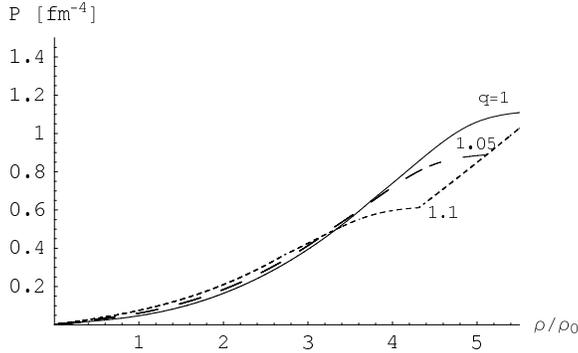}
} \caption{Pressure versus baryon density in units of the nuclear
saturation density $\rho_0$ in the mixed hadron-quark phase for
different values of $q$. Note that for $q=1.1$ the second transition
density, separating the mixed phase from the pure quark-gluon
matter, is reached.} \label{fig_MP1}
\end{figure}

\begin{figure}
\resizebox{0.55\textwidth}{!}{%
  \includegraphics*[90,550][500,800]{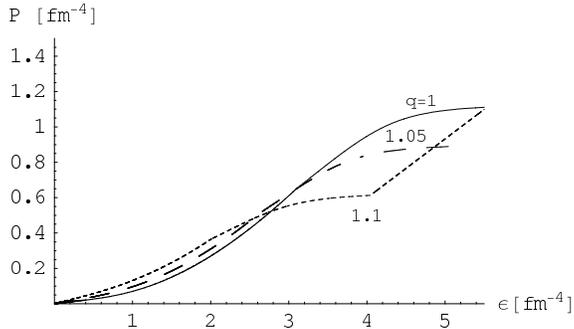}
} \caption{Pressure versus energy density in the mixed hadron-quark
phase for different values of $q$. As in the previous figure, we
note that for $q=1.1$ the second phase transition to a pure
quark-gluon matter is reached.} \label{fig_MP2}
\end{figure}

In Fig. \ref{fig_MP1}, we report the pressure versus baryon number
density (in unit of the nuclear saturation density $\rho_0$) and,
in Fig. \ref{fig_MP2}, the pressure as function of the energy
density in the mixed hadron-quark phase for different values of
$q$. For the hadronic phase we have used the so-called GM2 set of
parameters \cite{glen} and in the quark phase the bag parameter is
fixed to $B^{1/4}$=170 MeV. The temperature is fixed at $T=60$ MeV
and the proton fraction at $Z/A$=0.4, physical values which are
estimated to be realistic for high energy heavy-ion collisions.
The mixed hadron-quark phase starts at $\rho=3.75\,\rho_0$ for
$q=1$, at $\rho=3.31\,\rho_0$ for $q=1.05$ and at
$\rho=2.72\,\rho_0$ for $q=1.1$. It is important to observe that
for $q=1.1$ it is also reached the second critical transition
density, separating the mixed phase from the pure quark-gluon
matter phase, at $\rho=4.29\,\rho_0$ while for $q=1.05$ the second
critical density is reached at $\rho=5.0\,\rho_0$ and at
$\rho=5.57\,\rho_0$ for $q=1$.

As a concluding remark we note that non-extensive
statistical effects become extremely relevant at large baryon density
and energy density, as the ones which can be reached in high energy collisions experiments.
This fact can be an important ingredient in the
realization of a hydrodynamic model as well as to obtain a deeper microscopic
connection with the experimental observables.

\end{document}